\documentclass[notitlepage,twocolumn,superscriptaddress,showpacs,pre]{revtex4-1}

\usepackage{xcolor}
\usepackage{graphicx}
\usepackage{dcolumn}
\usepackage{bm}
\usepackage{amsmath}

\usepackage{latexsym}
\usepackage{float}
\usepackage{amssymb}
\usepackage{epsfig}
\usepackage{subfigure}
\usepackage{siunitx}
\usepackage{color}

\usepackage{cancel}
\usepackage{soul}

\begin{document}

\title{Direct Observation in 3d of Structural Crossover in Binary Hard Sphere Mixtures}

\author{Antonia Statt}
\affiliation{Institut f\"ur Physik, Johannes Gutenberg-Universit\"at Mainz,Staudinger Weg 9, 55128 Mainz, Germany}
\affiliation{ Graduate School of Excellence Materials Science in Mainz, Staudinger Weg 9, 55128 Mainz, Germany}
\author{Rattachai Pinchaipat}
\affiliation{H. H. Wills Physics Laboratory, University of Bristol, Tyndall Avenue, Bristol BS8 1TL, UK}
\author{ Francesco Turci}
\affiliation{H. H. Wills Physics Laboratory, University of Bristol, Tyndall Avenue, Bristol BS8 1TL, UK}
\author{ Robert Evans}	
\affiliation{H. H. Wills Physics Laboratory, University of Bristol, Tyndall Avenue, Bristol BS8 1TL, UK}	
\author{ C. Patrick Royall}
\affiliation{H. H. Wills Physics Laboratory, University of Bristol, Tyndall Avenue, Bristol BS8 1TL, UK}
\affiliation{ Centre for Nanoscience and Quantum Information, Tyndall Avenue, Bristol BS8 1FD, UK}
\affiliation{ School of Chemistry, University of Bristol, Bristol BS8 1TS, UK}

\date{\today}

\begin{abstract} 
For binary fluid mixtures of spherical particles in which the two species are sufficiently different in size, the dominant wavelength of oscillations of the pair correlation functions is predicted to change from roughly the diameter of the large species to that of the small species along a sharp crossover line in the phase diagram [C. Grodon, M. Dijkstra, R. Evans \& R. Roth, J.Chem.Phys. \textbf{121}, 7869 (2004)]. Using particle-resolved colloid experiments in 3d we demonstrate that crossover exists and that its location in the phase diagram is in quantitative agreement with the results of both theory and our Monte-Carlo simulations. In contrast with previous work [J. Baumgartl, R. Dullens, M. Dijkstra, R. Roth \& C. Bechinger, Phys.Rev.Lett. \textbf{98}, 198303 (2007)], where a correspondence was drawn between crossover and percolation of both species, in our 3d study we find that structural crossover is unrelated to percolation.
\end{abstract}

\pacs{82.70.Dd; 61.20.-p; 64.60.-i}        

\keywords{Suggested keywords}

\maketitle

\section{Introduction}
\label{sectionIntroduction}

Among the most striking occurrences in everyday life are phase transitions~\cite{cahn}. The familiar phase boundaries that delineate solids, liquids and gases are, of course, associated with non-analyticities of the free energy. However, there are other lines in the phase diagram where structural characteristics of the material change abruptly. For example, in one-component fluids the Fisher-Widom line distinguishes the region where the density-density (pair) correlation function exhibits damped oscillatory decay at large separations, characteristic of dense liquids, from that where the decay is monotonic, characteristic of gases and near critical fluids~\cite{fisher1969, evans1993}. In nature most materials are mixtures and increasing the number of components (species) leads to increasingly rich phase behaviour that is accompanied by further crossover lines. The most basic structural crossover in a binary mixture is predicted to occur when the dominant wavelength of oscillations in the pair correlation functions changes, from big to small, upon changing the composition~\cite{grodon2004,grodon2005}; this crossover is amenable to experimental investigation and is the subject of our study.

The binary hard sphere model is the key reference system for a simple liquid mixture, i.e. a mixture of two species of atoms or small molecules, since the structure of such mixtures is determined primarily by the repulsive forces acting between the atoms and these can be approximated by hard spheres~\cite{weeks1971}. At longer length scales, binary mixtures of colloidal particles immersed in a solvent can be prepared for which the effective colloidal interactions are, to an good approximation, hard sphere-like~\cite{royall2013}. Here we investigate such a colloidal system, using particle-resolved studies to extract pair-correlation functions~\cite{ivlev,royall2007jcp,thorneywork2014}, and test theoretical and simulation predictions~\cite{grodon2004,grodon2005} for fundamental features of the structure of binary mixtures.

We focus on a hard sphere mixture of big (\textit{b}) and small (\textit{s}) particles, characterized by the size ratio $q=\sigma_s/\sigma_b$, where $\sigma_b > \sigma_s$ denote the hard sphere diameters, and the packing (or volume) fractions $\eta_s$ and $\eta_b$. The existence of two length scales, associated with the two diameters, points to the possibility of physical phenomena associated with competition between these. In the supercooled liquid, this competition can be used to suppress crystalisation \cite{Zhang2014}. In the hard sphere crystalline state this competition gives rise to a wide variety of different phases, characterized by different crystal structures~\cite{bartlett1992,eldridge1993,hudson2011,Hopkins2011,filion2009}.

In the fluid state the competition leads to structural crossover revealed by considering the asymptotic decay, $r \rightarrow \infty $, of the primary structural indicator,  the pair correlation function $g_{ij}(r)$  between species $i$ and $j$. The $g_{ij}(r)$ are, of course, determined by the interaction potentials and a powerful means of analysing the nature of their decay is via a pole analysis of the Ornstein-Zernike (OZ) equations of liquid state theory \cite{hansen}.
We know from very general considerations of the mixture OZ equations in Fourier space~\cite{hansen,pearson1957}, that for short-ranged interparticle potentials the ultimate decay of all three pair correlation functions $g_{bb}(r), g_{bs}(r) $  and $g_{ss}(r)$ will be exponentially damped oscillatory with a common decay length $\alpha_0^{-1} \equiv \xi$ the true correlation length of the mixture and a common oscillatory wave length $2 \pi / \alpha_1$~\cite{evans1994}. Specifically we expect the total correlation functions $h_{ij}(r)=g_{ij}(r)-1$ to decay as
\begin{equation}
\label{eqh}
  r h_{ij}(r)\equiv r(g_{ij}(r)\!-\!1)) \! \sim \!A_{ij} e^{-\alpha_0 r}\cos{(\alpha_1 r \!-\!\theta_{ij})} 	
\end{equation}
for $r \rightarrow \infty$ with $i,j=b,s$. Only the amplitudes $A_{ij}$ and the phases $\theta_{ij}$  are species dependent and there are symmetry relations between these~\cite{evans1994}. Equation (\ref{eqh}) is a single pole approximation to the pair correlation functions: $\alpha_0$ and $\alpha_1$  are the imaginary and real parts of the leading order pole(s) of the partial structure factors $S_{ij}(k)$, determined as the complex root  $k=\alpha_1 + i \alpha_0$ of the equation
\begin{equation}\label{eqdk}
D(k)\equiv [1\!-\!\rho_s \hat c_{ss}(k)][1\!-\!\rho_b \hat c_{bb}(k)]\!-\!\rho_s\rho_b \hat c_{bs}(k)^2 =0
\end{equation}
having the smallest imaginary part $\alpha_0$. $D(k)$ is the common denominator entering the mixture OZ equations: $\hat c_{ij} (k)$ is the Fourier transform of the $ij$ pair direct correlation functions and $\rho_s$,$\rho_b$ are  the number densities of the two species~\cite{evans1994}.

Consider now a mixture at a large packing fraction  $\eta_b$ and small $\eta_s$. Intuitively, we expect the (common) wavelength of the oscillations to be approximately the diameter of the big species. On the other hand, when $\eta_b$ is small and $\eta_s$ is large the wavelength should be approximately the diameter of the smaller species. These observations are supported by theory and simulation~\cite{grodon2004,grodon2005}. Suppose now we prepare mixtures at different compositions.  What physics determines the crossover from oscillations with the wavelength $\sigma_b$ to those with wavelength $\sigma_s$?  Is there a sharp line, in the $\eta_s$ versus $\eta_b$ phase diagram, delineating a structural crossover whereby the wavelength of the longest ranged oscillations changes discontinuously?

Theory, based on a general pole analysis of the OZ equations, points to a \emph{sharp} structural crossover line and for (additive) hard spheres the line has been calculated explicitly within the Percus-Yevick (PY) and Density Functional Theory (DFT) approximations~\cite{grodon2004,grodon2005}. 
By sharp crossover we mean there is a line in the phase diagram where the wavelength of the slowest oscillatory decay of $g_{ij}(r)$ switches discontinuously from one value to another.
Although the $g_{ij}(r)$ obtained from Monte Carlo (MC) simulations of hard spheres in 3d~\cite{grodon2004} and of hard discs in 2d~\cite{grodon2005} are close to those from theory, simple visual inspection of $g_{ij} (r)$ is not sufficient to confirm a sharp crossover. In fact we shall show that demonstrating a sharp crossover requires determination of $\alpha_0$ and $\alpha_1$ in Eq. (\ref{eqh}) in order
to demonstrate a (sharp) change in the leading order pole controlling asymptotic behavior. While particle resolved experiments on mixtures have been performed previously \cite{ivlev,royall2007jcp}, obtaining the level of accuracy sufficient for the pole analysis is challenging \cite{ivlev,poon2012}. Here we show that it is possible to apply this pole analysis to experimental data and thereby provide compelling experimental evidence for a sharp crossover. We emphasize that the prediction of a sharp structural crossover transition is in no sense particular to hard spheres; it should be found for a very wide class of mixtures where the two species are of different sizes and the interparticle forces are short-ranged~\cite{archer2001}.

Significantly the same general argument for crossover pertains for a confined binary mixture, where one expects to see manifestations of structural crossover in the oscillatory (solvation) force arising from confinement between substrates \cite{grodon2005}, and for the oscillatory one-body density profiles at interfaces \cite{grodon2005,evans1994}. Specifically, the same (true) correlation length $1/\alpha_0$ and wavelength $2\pi/\alpha_1$ appearing in Eq. (\ref{eqh}) also determine the asymptotic decay of the solvation force and the density profiles. This connection has been addressed, in the context of confinement, in AFM experiments on one-component nanoparticles \cite{zeng2011, zeng2012} and, in the context of interfacial layering, for ionic liquids at sapphire substrates \cite{mezger2015}. We also note that the true correlation length is important in current research on the glass transition \cite{russo2015}.

A previous study, carried out by Baumgartl \emph{et al.}\;\cite{baumgartl2007}, on a binary mixture of small and big colloidal particles with size ratio $q=0.61$ provided some evidence for structural crossover and attempted to relate crossover to the sizes of networks containing connected big or connected small particles. However, Ref.\cite{baumgartl2007} could image only the first 2d colloidal layer next to the bottom wall of the sample and data analysis was performed from a 2d perspective. Moreover the pole analysis mentioned above was not employed so the authors were not able to ascertain 
the nature of crossover~\cite{baumgartl2007}. 
A later x-ray diffraction and microscopy  study on monolayers of bimodal nanoparticles reported packing structures resembling what might be expected for structural crossover~\cite{pontoni2009}.

In contrast, our present study employs 3d confocal microscopy combined with 3d MC simulations and liquid state theory for hard spheres to investigate the nature 
of the crossover. The paper is arranged as follows: in section \ref{sectionExperimental} we provide details of the experiments. Section \ref{sectionOneComponent} describes results for the asymptotic decay of $h(r)$ in a one-component colloidal system where we demonstrate that the inverse correlation length $\alpha_0$ and the wavelength $2\pi/\alpha_1$ can be determined from experiment, showing that a pole analysis of experimental and simulation data is feasible. In section \ref{sectionBinary} we present results for the binary system. Section \ref{sectionBinarySystemPairCorrelationFunctions}  describes $h_{ij}(r)$ as measured in experiment and in simulation. In section \ref{sectionBinaryPole} we fit both sets of data to a two-pole generalization of Eq. (\ref{eqh})  and show that both have the same pole structure as in PY approximation for hard sphere mixtures. It follows that our binary colloidal mixtures should exhibit the same sharp structural crossover. In section \ref{sectionPercolation}  we show that for the parameters of our experiments and simulations structural crossover is \emph{not} related to the size of networks of big and small particles, i.e. to percolation of one or other species.

\section{Experimental}
\label{sectionExperimental}

\begin{figure}
	\includegraphics[width=0.49\textwidth]{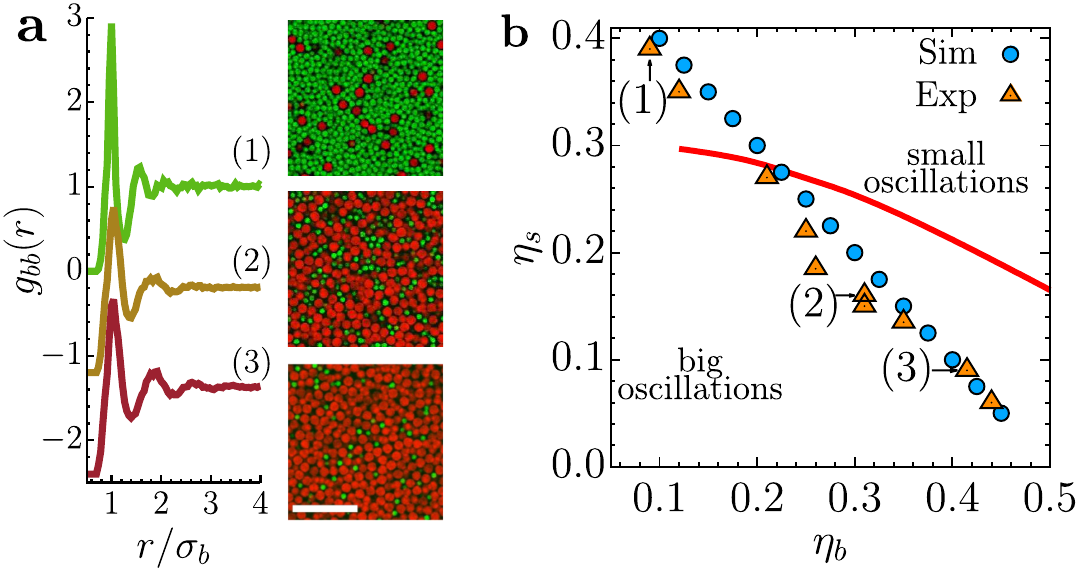}
	\caption{\label{figPhaseBinary} 
	    (a)
		Pair correlation functions $g_{bb}(r)$ measured for the big colloids at $\eta_s=0.39,\eta_b=0.09$ $(1)$,  $\eta_s=0.16,\eta_b=0.31$ $(2)$ and $\eta_s=0.09,\eta_b=0.415$  $(3)$. Data offfset for clarity. Corresponding confocal images of 	
		the mixture of red ($\SI{2.9}{\micro \metre}$) and green ($\SI{1.88}{\micro \metre}$) colloids are shown. The scale bar indicates  \SI{19}{\micro \metre}. 
		(b)
		 Packing fractions in the $(\eta_b,\eta_s)$-plane at which experiments (triangles) and hard sphere simulations (circles) are performed. The size ratio $q=0.648$ and the total packing fraction $\eta_\text{tot}=\eta_s+\eta_b$ is roughly $0.5$. The line is the crossover-line calculated from PY approximation which terminates at $\eta_b \approx 0.12$ \cite{grodon2004}. Above this line the $g_{ij}(r)$ show oscillations with wavelength $\sim \sigma_s$, and below the wavelength is $\sim\sigma_b$.
		We expect to find crossover on the experimental and simulation paths near their intersection with the PY crossover-line.
	}
\end{figure}

We used a Leica SP5 confocal microscope fitted with a resonant scanner. The colloids were suspended in a solvent mixture chosen to match the density and refractive index of the polymethyl methacrylate particles. In order to screen any residual electrostatic interactions, tetra-butylammonium bromide salt was added to the cyclohexyl bromide-$cis$-decalin solvent. 
A borosilicate glass square capillary with internal dimensions of \mbox{$0.50 \times \SI{0.50}{\milli \metre}$} and glass thickness of \SI{0.10}{\milli \metre} was filled with the suspensions and sealed at each end with epoxy glue to prevent evaporation.

In our study of the pair correlation function of a one-component system three different sizes of PMMA particles were used. One set had a diameter $\sigma$ of \SI{3.23}{\micro \metre} and 6\% polydispersity (determined with SEM). Samples were prepared at $\eta=0.587$ and $\eta=0.593$. The second set, with \SI{2.9}{\micro \metre} diameter and 5\%  polydispersity were prepared at the packing fraction $\eta=0.561$. Samples with the lowest packing fractions, $\eta=0.496$ and $\eta=0.365$, were prepared using particles of size \SI{1.88}{\micro \metre} and  5\%  polydispersity.

Binary mixtures of polymethylmethacrylate (PMMA) particles of diameter \SI{2.9}{\micro \metre} and \SI{1.88}{\micro \metre} were prepared at different densities as shown in Fig.\ref{figPhaseBinary}(b). The particle size and polydispersity were determined using static light scattering and the polydispersity was  5\%. We note that the effects of polydispersity were examined in the DFT study by Grodon et al. who found that for a rather broad bimodal distribution of diameters a clear signature of crossover was present -- see Figs. 8,9 of Ref.~\cite{grodon2005}. For a binary mixture the coordinates of the two different particle species are tracked separately and the overlaps between the two different particle types are removed afterwards. From the coordinates of the colloids, the pair correlation functions $g_{ij}(r)$ can be determined. An example is shown for $g_{bb}(r)$ in Fig. \ref{figPhaseBinary}(a).

\section{Asymptotic decay in the one-component system}
\label{sectionOneComponent}

\begin{figure*}
	\centering
	\includegraphics{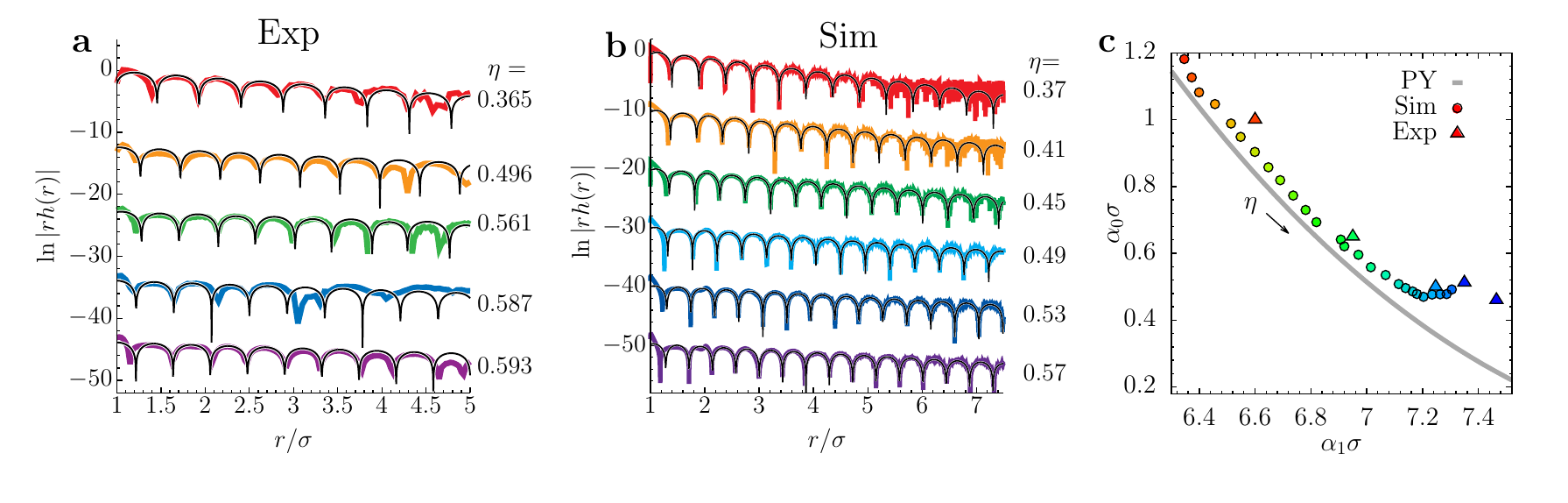}
	\caption{\label{figGOne}
		\textbf{(a)}
		Plots of $\ln{|rh(r)|}$for a one component colloidal liquid 
		for several packing fractions $\eta$ as indicated: \textbf{(a)} experiment \textbf{(b)} simulations of the hard sphere fluid. The curves are shifted vertically for clarity.
		The black lines are fits to Eq. (\ref{eqhcos}) for the total correlation function $h(r)$, from which $\alpha_0$ and $\alpha_1$ are determined. 
		\textbf{(c)} Comparison of the values of $\alpha_0$ and $\alpha_1$ calculated from the experimental 
		results (triangles) and from simulations (circles) and the PY approximation (line) for
		the one component hard sphere fluid. The experimental results pertain to the five values of $\eta$ given in (a). Simulation results cover the range from red $\eta=0.37$
		to blue $\eta = 0.57$. 	
		The PY result corresponds to the leading order pole for the hard sphere fluid, plotted for $\eta=0.35$ to $\eta=0.62$. 
		}
\end{figure*}

As it can be challenging 
to determine directly the packing fraction $\eta$ in experiments~\cite{poon2012,royall2013}, we fit the Fourier transform of the hard sphere structure factor, given by the PY approximation~\cite{hansen}, to the first two maxima in $g(r)$ as obtained from the confocal microscopy measurements. This procedure yields the values of $\eta$ given in Fig. \ref{figGOne}(a) \cite{footnoteEquivalent}.
Experimental results for the decay of $h(r)$ 
are shown in Fig.\ref{figGOne}(a) where the black lines are fits of this data to the equation 
\begin{align}\label{eqhcos}
rh(r) \sim A e^{-\alpha_0 r}\cos{(\alpha_1r -\theta)}, \quad r \rightarrow \infty 
\end{align}
which is the one-component version of Eq.(1). In this equation $\alpha_0$ and $\alpha_1$  are the imaginary and real parts of the leading order pole of the structure factor. The poles are given by the complex roots of  
\begin{align}\label{eqck}
1-\rho \hat c(k) =0 
\end{align}
where $\hat c(k)$ is the Fourier transform of the pair direct correlation function at number density $\rho$. There is an infinite number of poles
and the leading order pole $k=\alpha_1 + i\alpha_0$ is that with the smallest imaginary part $\alpha_0$, closest to the real axis. Equation (\ref{eqhcos}) is a single-pole approximation that retains only the slowest decaying contribution to $h(r)$~\cite{evans1993,evans1994,hansen}. Fits were performed over the region $1.5 < r/\sigma < 4.0$ as the data becomes noisier at larger distances $r$. We see that the fits are rather good and allow us to determine reliable values for $\alpha_0$, the inverse of the true correlation length, and for the wavelength $2 \pi/\alpha_1$. 
The resulting values of $\alpha_0$ and $\alpha_1$ are plotted, as triangles, in Fig.\ref{figGOne}(c) for five values of $\eta$. As $\eta$ increases, $\alpha_0$ decreases and $\alpha_1$ increases until we reach a value of $\eta  \sim 0.54 $ beyond which $\alpha_0$ appears to stay roughly constant while $\alpha_1$ continues to increase. These results show that it is possible to extract accurate $\alpha_0$ and $\alpha_1$ and thus to apply the pole analysis to experimental data.

In Fig.\ref{figGOne}(b) we present results for $\ln{ |rh(r)|}$ from MC simulations of a one-component hard sphere liquid. These were carried out in the NVT ensemble in a cubic box of side roughly $20 \sigma$.
Recall that pure hard spheres undergo an equilibrium freezing transition to an fcc crystal at $\eta =0.492$~\cite{frenkel}. Thus for several of the packing fractions we consider, and in particular for $\eta$  larger than about $0.54$, the hard sphere fluid may start to crystallise during the simulation. We took care that any runs which did indeed crystallise were excluded from our analysis, by using the averaged bond orientational order parameters $\bar q_4$ and $\bar q_6$ to identify crystallisation \cite{lechner2008}.  It follows that the range of simulation times, and also 
the maximum packing fraction, are limited. The black lines are a fit to Eq.(\ref{eqhcos}), now over the larger range $1.5 < r/\sigma < 5.5$. The fit is excellent for all values of $\eta$ investigated and the resulting values for $\alpha_0$  and $\alpha_1$ are plotted as circles in Fig.\ref{figGOne}(c). These follow closely the trend of the experimental results and show $\alpha_0$ flattening off for the largest values of $\eta$ considered.

It is important to recall that Eq.(\ref{eqhcos}) is valid for the {\it{long}} range decay of $h(r)$ and therefore we cannot expect that the first two maxima are well-matched by this single pole approximation. The plots in Fig.\ref{figGOne}(a)  show this to be the case. However, as predicted by the early theoretical work, Eq.(\ref{eqhcos}) provides a remarkably good fit for both experimental and simulation data at {\it{intermediate}} distances, i.e. separations $r$ as low as second nearest neighbours. As we shall see below, the same conclusion holds for the single pole approximation Eq. (\ref{eqh}) pertinent to binary mixtures, provided the state point is away from crossover. The line in Fig. \ref{figGOne}(c) corresponds to $\alpha_0$  and  $\alpha_1$ calculated from the leading order pole of the hard sphere structure factor obtained from  PY theory, i.e. the solution of Eq.(\ref{eqck}) where $\hat c (k)$ is the Fourier transform of the PY direct correlation function~\cite{grodon2004,evans1993,evans1994}. We see that the simulation results lie slightly above the PY results for all values of $\eta$, with the difference becoming more pronounced at high values. As expected, the PY results for $\alpha_0$ continue to decrease monotonically at large values of $\eta$, implying the (true) correlation length continues to increase. There is no indication of the flattening off for $\eta> 0.54$ that is observed in simulation. Recall that the PY approximation is not especially accurate at large values of $\eta$. For example the PY compressibility equation of state for hard spheres yields a pressure that is already significantly larger that the simulation result for values of $\eta$ slightly below freezing~\cite{hansen}.

\begin{figure*}
	\centering
	\includegraphics[width=17cm]{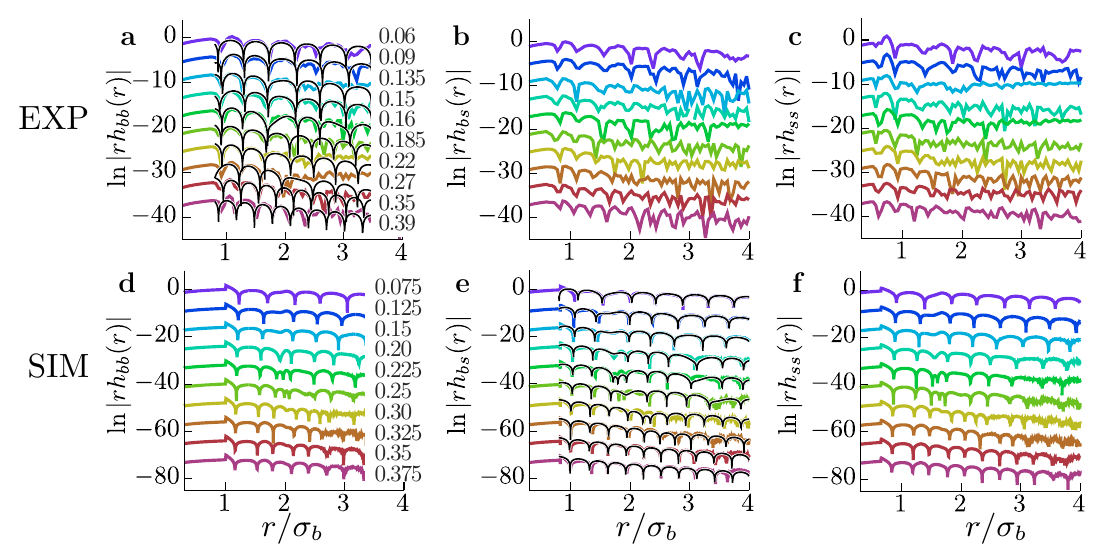}
	\caption{\label{figGBinary} 
	 Plots of  $\ln|r h_{ij}(r)|$ obtained by experiment (top) and simulation (bottom) for big-big, big-small, and small-small total correlation functions. The packing fraction $\eta_s$ (marked) is increasing from top (blue) to bottom (red) for each case. Curves are shifted vertically for clarity. The black lines in (a) and (e) are fits from which  $\alpha_0$, $\alpha'_0$, $\alpha_1$, and $\alpha'_1$ in Eq.(\ref{eqhbinary}) are obtained. The simulations correspond to size ratio $q=0.648$.}
\end{figure*}

\section{Asymptotic decay in the binary system}
\label{sectionBinary}

\subsection{Binary system pair correlation functions} 
\label{sectionBinarySystemPairCorrelationFunctions}

We now consider the case of binary hard spheres for which crossover occurs. In Fig. \ref{figPhaseBinary}(a) $g_{bb}(r)$, denoting $bb$ or red-red particle correlations, is plotted alongside typical confocal images, for three compositions. We prepared each sample at a total packing fraction $\eta_\text{tot} \approx 0.5 $ to ensure that the oscillations in $g_{ij}(r)$ decay sufficiently slowly so that 4-5 oscillations can be observed. The packing fractions $\eta_b$ and $\eta_s$  are determined by fitting to results for the hard sphere $g_{ij}(r)$. \cite{footnoteEquivalent}.  Figure \ref{figPhaseBinary}(b) displays the 10 state points (triangles) in the $\eta_s$ versus $\eta_b$ plane.

Results for  $\ln |r h_{ij}(r)|$  are plotted in the top row of Fig. \ref{figGBinary}. The bottom row shows corresponding results of simulations  with the experimental size ratio $q=0.648$, selected from the state points (circles) shown in Fig.\ref{figPhaseBinary}(b). The total packing fraction is kept constant at $\eta_\text{tot}=0.5$ for all simulations, resulting in particle numbers between $N_b=6875 $ and $N_b=1527$ and $N_s=2807$ and $N_s=22460$, for the big and small particles, respectively, in a simulation volume with periodic boundaries of size roughly $20 \sigma_b \times 20 \sigma_b \times 20 \sigma_b$. We observe that for all three $h_{ij}(r)$, in both experiment and simulation, the wavelength of the (damped) oscillations is roughly $\sigma_b$ for the blue curves (rich in $b$) and roughly $\sigma_s$ for the red curves (rich in $s$). At intermediate values of $\eta_s$ there is interference between the two length scales and Eq.(\ref{eqh}) is no longer sufficient to describe the observed behaviour. Crossover clearly occurs but from visual inspection it is dificult to identify whether it is sharp or not. Nor can we determine the value of $\eta_s$ where it occurs. A more powerful method of analysis is required.

\subsection{Pole analysis}
\label{sectionPoleAnalysis}
\label{sectionBinaryPole}

\begin{figure}
	\centering
	\includegraphics[width=8cm]{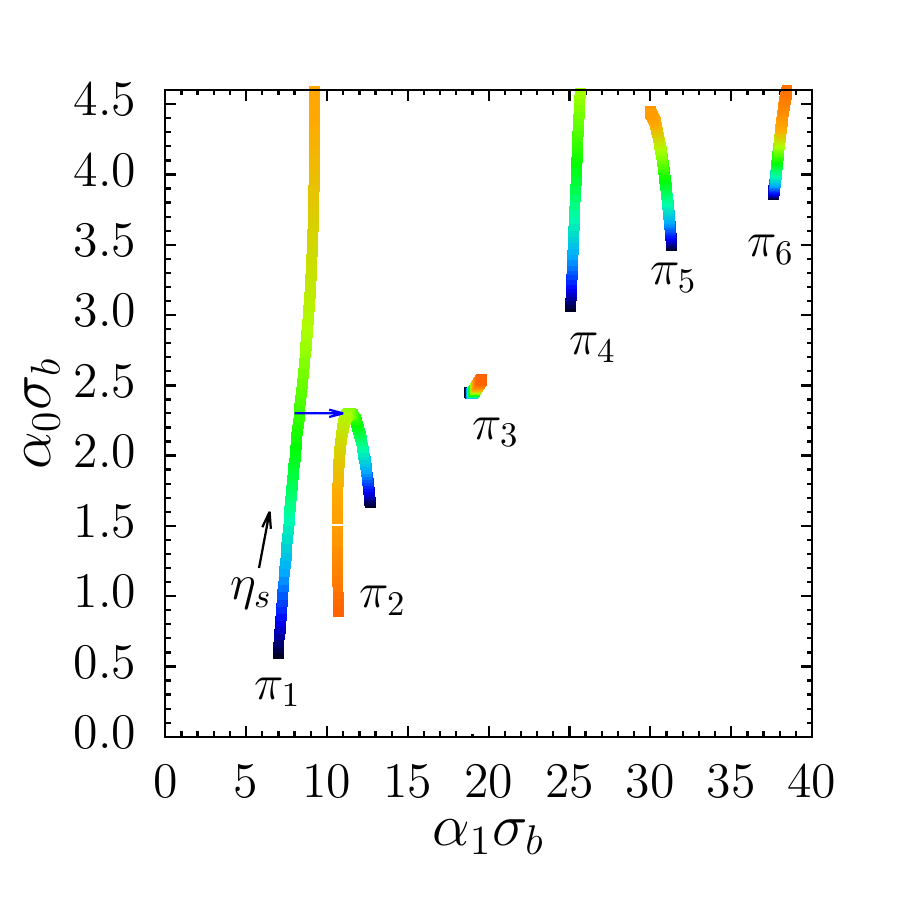}
	\caption{\label{figGenesis} Pole structure of a two component hard sphere fluid treated
		in the PY approximation for size ratio $q=0.648$ and fixed $\eta_s+\eta_b=0.5$. 
		The colour indicates the value of the packing fraction $\eta_s$ increasing from blue $\eta_s=0.01$
		to red $\eta_s = 0.49$.  Crossover occurs at  $\eta_s=0.28$ where $\alpha_0\sigma_b=2.3$, as indicated by the horizontal arrow. 
	}
\end{figure}


Having demonstrated in section \ref{sectionOneComponent} that the pole analysis can be applied to experimental data, we now apply it to the binary system and elucidate the origin of the structural crossover.
In Fig. \ref{figGenesis} we plot the real and imaginary parts of the six lowest lying poles (those closest to the real axis) as a function of increasing $\eta_s$ for a binary hard sphere mixture with $q =0.648$ at fixed $\eta_s+\eta_s =0.5$ treated within PY approximation~\cite{grodon2004,grodon2005,evans1994}, which represents the path through the phase diagram (Fig. \ref{figPhaseBinary}(b)) taken in the simulations and, approximately, that of the experiments.
The poles lie on separate branches which are labelled $\pi_i$ in Fig.4. The two branches $\pi_1$ and $\pi_2$ ,where the imaginary part $\alpha_0$ is smallest, determine the leading order decay of the pair correlation functions [4, 5] and Equation (\ref{eqhbinary}) below includes contributions from these two branches. At small values of $\eta_s$ (colour blue) the imaginary part $\alpha_0$ is smaller on the $\pi_1$  branch than on the $\pi_2$ branch so the dominant decay of the total correlation functions $h_{ij}(r)$ has oscillations with wavelength $2 \pi / \alpha_1$  corresponding to roughly the diameter of the big spheres. On the other hand, at large values of $\eta_s$ (colour red) $\alpha'_0$  on the $\pi_2$  branch is smaller so the dominant decay of oscillations has wavelength $2 \pi / \alpha'_1$  corresponding to roughly the diameter of the small spheres. At some intermediate value of $\eta_s$ there is a sharp crossover whereby the decay at large $r$ of the three pair correlation functions switches from being governed by the branch $\pi_1$ to being governed by the branch $\pi_2$.  For a sharp crossover to occur there must be separate branches $\pi_1$ and $\pi_2$ in the $(\alpha_1, \alpha_0)$ plane. If there is only one lowest-lying pole, as in the one component case, sharp crossover cannot occur. For this particular system the crossover occurs, as indicated by the horizontal arrow, at a value of {$ \eta_s=0.28$}. The higher order poles $\pi_3$  to $\pi_6$  shown in Fig. \ref{figGenesis} play no role in the crossover.

\begin{figure}
	\centering
	\includegraphics[width=8cm]{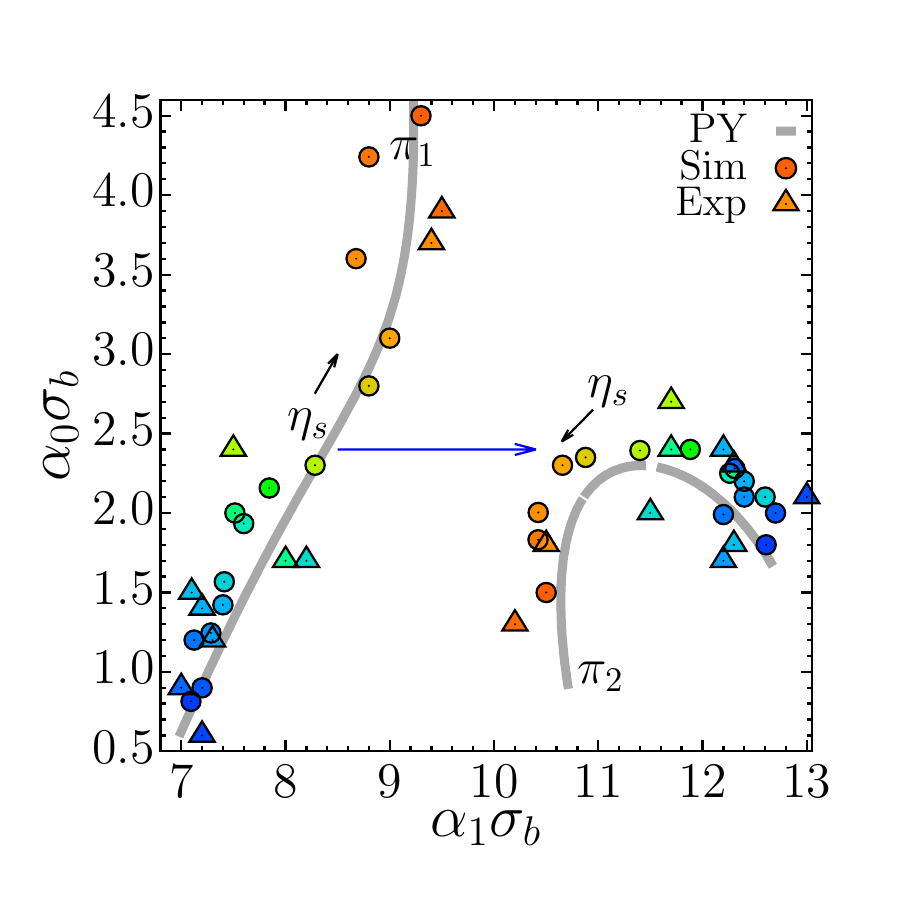}
	\caption{\label{figPoleBinary} 
	Pole structure of leading two poles as the composition is varied for $\eta_b+\eta_s\approx0.5$.
	Circles: results of fits with functions of the form Eq.(\ref{eqhbinary}) to $\ln|r h_{bs}(r)|$ from simulations of hard sphere mixtures, where $\eta_s$ is varied from $0.03$ (blue) to $0.55$ (red). Triangles: corresponding fits to $\ln|r h_{bb}(r)|$ from experiment. The values of ($\eta_b,\eta_s)$ are given in Fig.\ref{figPhaseBinary}(b). Crossover occurs at the statepoint ($\eta_b,\eta_s$) where the values of $\alpha_0$ in each branch are equal, indicated by the horizontal arrow. For $q=0.648$ this is close to (0.22,0.28) and $\alpha_0\sigma_b=2.3$ (PY), $\alpha_0\sigma_b\approx 2.4$ (Sim), $\alpha_0\sigma_b\approx 2.5$ (Exp). The grey lines show the PY approximation for the first pole $\pi_1$ and  second pole $\pi_2$. 
	 }
\end{figure}

\begin{figure}
	\centering
	\includegraphics[width=8cm]{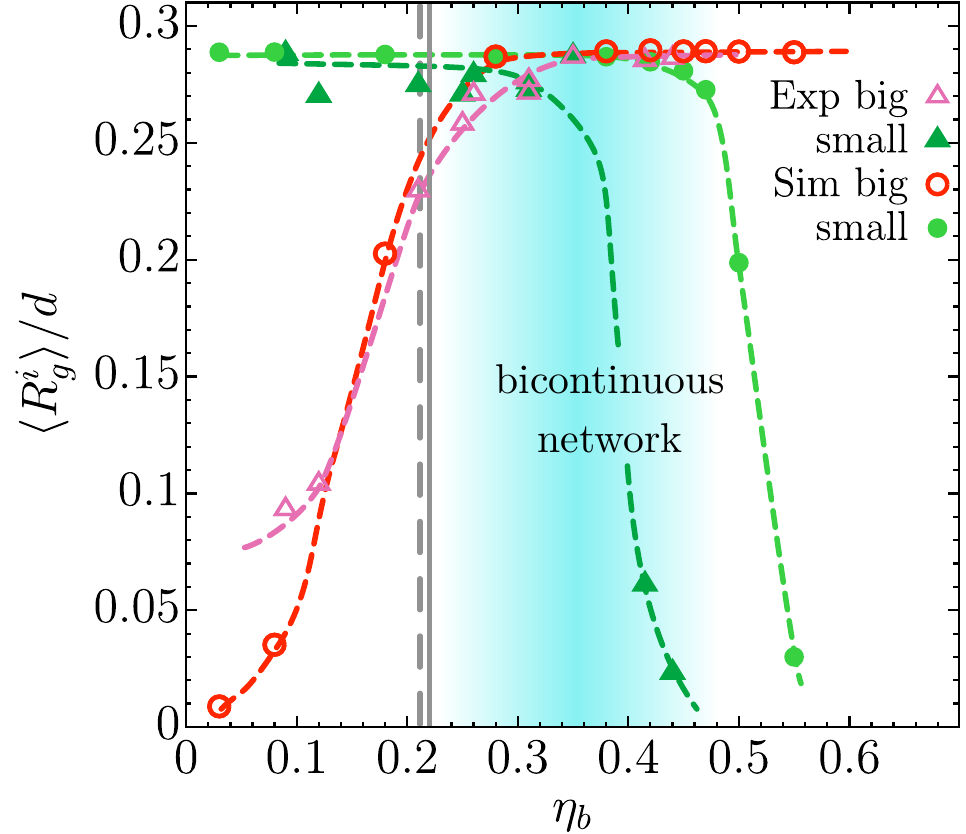}
	\caption{\label{figPercBinary}
     	The weighted {average} of the radius of gyration $\langle R_g^i \rangle$ of species $i$, divided by the box diagonal $d$ as a function of packing fraction $\eta_b$ of the big particles. 
     	A percolating cluster that spans the system has $\langle R_g^i \rangle/d=1/(2\sqrt{3})=0.289$.
     	Results  from simulation (circles) and experiment (triangles) are shown. The values of  $(\eta_b,\eta_s)$ are given in Fig. \ref{figPhaseBinary}(b). 
     	The vertical solid line indicates the location of the crossover line as inferred from PY and simulations while the vertical dashed line shows the estimate from experiment.  }
\end{figure}

Figure \ref{figGenesis} shows, based on PY theory, that we may expect the crossover to be dominated by the leading order poles $\pi_1$  and $\pi_2$. We now seek to identify these poles in the experiments and simulations. We therefore fit the pair correlation functions with the expression 
\begin{align}\label{eqhbinary}
      r h_{ij}(r) &\sim A_{ij}e^{-\alpha_0 r} \cos{(\alpha_1 r-\theta_{ij})} \nonumber \\  &+  A'_{ij}e^{-\alpha'_0 r} \cos{(\alpha'_1 r-\theta'_{ij})}, \quad r \rightarrow \infty 
\end{align}
which corresponds to the two-pole approximation~\cite{grodon2004}. Examples of such fits, denoted by the black lines, are shown in Fig. \ref{figGBinary}(a) $h_{bb}(r) $ (experiment) and \ref{figGBinary}(e) for $h_{bs}(r)$ (simulation). Fits are performed in the region $1.8 < r/\sigma_b< 4.0 $. For larger separations statistics are limited and for smaller separations the approximation Eq.(\ref{eqhbinary}) is no longer appropriate. The fits determine the first pole: $\pi_1=\alpha_1+ i\alpha_0$, and the second: $\pi_2=\alpha'_1 + i\alpha'_0$. Which pole contribution dominates at longest range in Eq.(\ref{eqhbinary}) depends on whether $\alpha_0$ is greater or smaller than $\alpha'_0$ and  on the composition~\cite{grodon2004}. In Fig. \ref{figPoleBinary} we plot the resulting values of the four parameters for all the compositions investigated. The results indicate two separate branches $\pi_1$ and $\pi_2$ ,plotted for experiments(triangles) and simulations (circles). Remarkably both sets of results lie rather close to those from PY theory for the binary hard sphere mixture, with the same size ratio (grey lines), obtained by solving Eq.(\ref{eqdk}) with the PY direct correlation functions. 
For the present system crossover occurs at $\eta_b=0.22$ (PY), $\eta_b = 0.21(4)$ (Exp) and $\eta_b = 0.22(2)$ (Sim). Figure \ref{figPoleBinary} thus provides strong evidence that the experimental system follows closely the predictions of theory and simulation, showing that in the binary colloidal mixture there is 
a sharp crossover.

\subsection{Percolation Analysis}
\label{sectionPercolationAnalysis}
\label{sectionPercolation}

We now consider the relationship, if any, between crossover and connectivity of each species. In Ref.~\cite{baumgartl2007} the authors performed an analysis of real-space configurations by calculating the extent of networks corresponding to $bb$ or $ss$ nearest neighbour bonds based on a Voronoi construction. They argued percolation was related to crossover as in their 2d approach both occurred around the same state point. We follow their procedure and consider the radius of gyration of a network of regions of each species: 
\begin{align}
R_g^i=\sqrt{\frac{1}{n^i} \sum\limits_{k=1}^{n^i}({\bf{x}}_k^i-{\bf{R}}_0^i)^2},
\end{align}
with ${\bf{R}}_0^i$ the center of the network. Assuming there are $N_C^i$ networks formed by  
$n^i$ connected particles of species $i$, we calculate a weighted average of the radii of gyration 
\begin{align}
\langle R_g^i \rangle = \frac{1}{N^i}\sum\limits_{m=1}^{N^i_C}n^i(m)R^i_g(m),
\end{align}
where $N^i$ denotes the total number of particles of species $i$. In analysing the experimental and simulated configurations, a Voronoi tessellation~\cite{rycroft2007,rycroft2006} of the networks of big and small  particles was performed. The Voronoi tessellation provides a list of neighbours for each particle in the sample. The experimental data has no periodic boundary conditions, therefore the edges were cut ($10$ pixels in each dimension) after the Voronoi tessellation and the networks inside the sample  evaluated. Two particles which share a face of their respective Voronoi polyhedra are deemed connected. In the list of neighbours the clusters (in networks of the big and small particles) need to be identified, for which we use a depth first search~\cite{cormen2001}.  The weighted average of the radii of gyration $\langle R_g^i \rangle$ is shown in Fig.\ref{figPercBinary} as a function of $\eta_b$ for both experiment and simulation \footnote{We choose to divide the averaged radius of gyration by the box diagonal $d$. Although the simulation box is cubic, the experimentally measured sample volume is not. Each  has a length of $512$ pixels of size $178$ nm ($0.062$ $\sigma_b$) in $x$- and $y$-direction, whereas the $z$-direction is smaller and varies. Typical images depths are $100$--$200$ pixels of size $168$ nm ($0.058$ $\sigma_b$).}.

Figure~\ref{figPercBinary} shows the onset of percolation for species $b$ and $s$ in both simulation and experiment. The percolation behaviour of the big particles is very similar in experiment and simulation. 
In both cases structural crossover, inferred from our pole analysis of $h_{ij}(r)$ and marked by a vertical line (solid for PY theory and simulation, dashed for the fits to the experimental data), occurs prior to the percolation threshold for species $b$. For the small particles percolation has already occurred for values of $\eta_b$ much larger than the crossover value. On the basis of these results, it is difficult to see how structural crossover can be linked to onset of percolation. This conclusion differs from that of Ref.~\cite{baumgartl2007} whose analysis is based on 2d systems where there is a single point for percolation. We suggest that it could be a coincidence that percolation occurred close to crossover in the dense mixtures of Ref.~\cite{baumgartl2007}. Our present results show that crossover is not related to percolation in 3d. Furthermore we know that  crossover persists to extreme dilutions for hard sphere mixtures, at least for size ratio $q=0.5$ \cite{grodon2005}, whereas percolation cannot occur in a dilute system. Moreover crossover was found in the exactly solvable model of a binary hard-rod mixture in 1d  \cite{grodon2005} and there is no percolation in 1d.

\section{Discussion and Conclusions} 
\label{sectionDiscussion}

We investigated a fundamental aspect of the structure of bulk liquid mixtures, namely the decay of the pair correlation functions $g_{ij} (r)$. For binary hard sphere mixtures, Percus-Yevick theory and density functional theory predict that the  dominant wavelength of the  oscillations in $g_{ij} (r)$ should change abruptly at a sharp crossover line in the $\eta_b$ versus $\eta_s$ phase diagram~\cite{grodon2004,grodon2005}.  The pair correlation functions $g_{bb}(r),g_{bs}(r)$, and $g_{ss}(r)$ obtained in our particle-resolved experiments and Monte Carlo simulations exhibit clear structural crossover, \emph{i.e.} the wavelength of the oscillatory decay changes from approximately the diameter of the large particles to the diameter of the small particles as the relative amount of small particles is increased. 

In order to investigate the nature of the crossover, i.e. whether or not it is sharp, we have shown that it is possible to apply the pole analysis to experimental data and this provides strong evidence that crossover is indeed sharp. For the size ratio and total packing fraction we consider, fitting the functional form Eq.(\ref{eqhbinary}), which allows for the presence of two wavelengths, to the experimental and simulation data provides compelling evidence for pole structure similar to that from theory -- see Fig.\ref{figPoleBinary}. 
Moreover this enables us to locate the crossover point in the phase diagram and we find it is very close to the results of theory and simulation. Note that although the theory is based strictly on asymptotic analysis of the mixture Ornstein-Zernike equations, our experiments and simulations show that the predictions remain valid for the intermediate range decay of $g_{ij} (r)$, \emph{i.e.} for $r>$ second nearest neighbor. Note also that our colloidal particles exhibit a degree of polydispersity whereas the present theory and simulations take no account of this.

We observe that the experiments and simulations, performed at $\eta_\text{tot} =0.5$, display bicontinuous percolation over a range of $\eta_b$. That crossover does not occur within this range reinforces our argument that percolation and crossover are, in general, unrelated phenomena.

Finally we emphasize that structural crossover is not particular to binary hard spheres. We can expect similar behaviour for many dense binary liquid mixtures, such as metals, Noble gases and molecules such as CCl$_4$ and globular proteins which may reasonably be treated as spherical. Neutron scattering experiments, in particular those which enable the oscillatory decay of $g_{ij} (r)$ to be extracted from the partial structure factors \cite{salmon2006, salmon2013}, could elucidate further the crossover in a wide range of materials. 

\begin{acknowledgments}
AS acknowledges the Graduate School of Excellence Materials Science in Mainz, Staudinger Weg 9, 55128 Mainz, Germany for financial support and the University of Bristol for hospitality. 
CPR acknowledges the Royal Society and Kyoto University SPIRITS fund for financial support and FT and CPR acknowledge the European Research Council (ERC consolidator grant NANOPRS, project number 617266). 
EPSRC grant code EP/ H022333/1 is acknowledged for provision of the confocal microscope used in this work. 
RP acknowledges the Development and Promotion of Science and Technology Talents Project for a Royal Thai Scholarship.
\end{acknowledgments}

\end{document}